\documentclass[aps,prl,twocolumn,superscriptaddress]{revtex4-1}  

\usepackage{graphics}  
\usepackage{dcolumn}   
\usepackage{bm}        
\usepackage{amssymb}  
\usepackage{amsmath}

\usepackage{xcolor}
\usepackage{url}

\begin{document}
\widetext
\title{Determination of quantum defect for the Rydberg P series of Ca~II}

\author{Arezoo Mokhberi}
\affiliation{QUANTUM, Institut f\"ur Physik, Johannes Gutenberg-Universit\"at Mainz, Staudingerweg 7, 55128 Mainz, Germany}
\author{Jonas Vogel}
\affiliation{QUANTUM, Institut f\"ur Physik, Johannes Gutenberg-Universit\"at Mainz, Staudingerweg 7, 55128 Mainz, Germany}
\author{Justas Andrijauskas}
\affiliation{QUANTUM, Institut f\"ur Physik, Johannes Gutenberg-Universit\"at Mainz, Staudingerweg 7, 55128 Mainz, Germany}
\affiliation{Helmholtz-Institut Mainz, Staudingerweg 18, 55128 Mainz, Germany}
\author{Patrick Bachor}
\affiliation{QUANTUM, Institut f\"ur Physik, Johannes Gutenberg-Universit\"at Mainz, Staudingerweg 7, 55128 Mainz, Germany}
\affiliation{Helmholtz-Institut Mainz, Staudingerweg 18, 55128 Mainz, Germany}
\author{Jochen Walz}
\affiliation{QUANTUM, Institut f\"ur Physik, Johannes Gutenberg-Universit\"at Mainz, Staudingerweg 7, 55128 Mainz, Germany}
\affiliation{Helmholtz-Institut Mainz, Staudingerweg 18, 55128 Mainz, Germany}
\author{Ferdinand Schmidt-Kaler} 
\affiliation{QUANTUM, Institut f\"ur Physik, Johannes Gutenberg-Universit\"at Mainz, Staudingerweg 7, 55128 Mainz, Germany}

\date{\today}


\begin{abstract}
We present an experimental investigation of the Rydberg 23\,P$_{1/2}$ state of single, laser-cooled $^{40}$Ca$^+$ ions in a radiofrequency ion trap.
Using micromotion sideband spectroscopy on a narrow quadrupole transition, the oscillating electric field at the ion position was precisely characterised, and  the modulation of the Ryd- berg transition due to this field was minimised.
From a correlated fit to this P line and previously measured P and F level energies of Ca~II, we have determined the ionization energy of 95\,751.916(32) $\rm {cm}^{-1}$, in agreement with the accepted value, and the quantum defect for the $n$\,P$_{1/2}$ states.
\end{abstract}

\maketitle

\section{Introduction}\label{sec:Int}
Spectroscopy of ionic Rydberg states yields experimental data that are essential for determining various properties of Rydberg ions. 
The necessity of such data stems from the fact that quantum defect theory \citep{seaton83a} (which provides a theoretical basis for predicting properties of high-lying states for any charge of a core) does not describe the nontrivial behaviour of core-penetrating Rydberg states with low angular momentum $l<4$ for hydrogen-like atoms and ions.
For many applications, modifications to the purely Coulomb part of the interaction can be represented by a single, weakly energy-dependent phase shift called quantum defect that has to be experimentally determined \citep{gallagher05a}. 

Rydberg states of atoms, exhibit remarkable properties, such as high sensitivity to external electric and magnetic fields. 
The key advantage is the largely enhanced polarisability, which make these systems ideal for fundamental cavity quantum electrodynamics experiments~\citep{raimond01a, brune96a} and for precise measurements as a quantum sensor of electric fields with unmatched sensitivity~\citep{facon16a, brune94a}.
For trapped Rydberg ions, strong dipole-dipole interactions can be achieved by microwave dressing fields~\citep{mueller08a}. 
Another exciting feature of Rydberg ions confined in a trap is the coupling between their external and internal degrees of freedom owing to their large polarisabilities. Therefore they are excellent platforms for exploring non-equilibrium dynamics in the quantum regime in an extremely controllable fashion \citep{li12a, silvi16a} as well as for extending quantum simulation techniques from Rydberg atoms~\citep{saffman10a} to ionic species.

Experimental investigations of Rydberg states of trapped ions, inspired by reference~\citep{mueller08a}, have been commenced with the first realisation of Rydberg F~states of Ca$^{+}$ ions \citep{feldker15a}. Further development have been recently made in a seminal experiment for coherent control of a single trapped Sr$^{+}$ ion in Rydberg S~states~\citep{higgins17b}. 
However trapped ions have been widely explored over the past three decades for exciting applications, such as quantum information processing~\citep{haeffner08a}, their Rydberg properties are not well understood. An important development is envisioned by implementing fast quantum gates using Rydberg ions~\citep{li14a, vogel19a}.

Rydberg-state spectroscopy in combination with Rydberg-series extrapolation presents the most precise method for determining the ionization energy \citep{herzberg72a, rydberg1890a, ritz1908a}, which is an important quantity for experimental studies and serves as a reference data for testing {\it ab initio} calculations. 
Despite the ubiquitous use of Ca$^{+}$ ions in quantum technology experiments, up to date, there is no precise measurement of its ionization energy.
 
Here, we report on Rydberg excitation of single, laser-cooled Ca$^{+}$ ions in a radiofrequency (RF) ion trap from the metastable 3\,D$_{3/2}$ state to the 23\,P$_{1/2}$ state. Measurements of the oscillating electric field at the ion position and its influence on the Rydberg line shape are discussed (sections~\ref{sec:Exp} and \ref{sec:Line}). This observation in combination with previously measured data on the P and F levels of Ca$^{+}$~\citep{NISTasd, feldker15a} enabled us to determine the quantum defect of the P and F levels as well as the ionization energy of Ca II (section~\ref{sec:Qdef}).

\section{Experimental setup for Rydberg excitation of cold trapped ions}\label{sec:Exp}
To confine ions, we employed a linear RF ion trap, consisting of four gold-coated, blade electrodes and two titanium endcaps. This blade trap features 750 $\mu$m ion-electrode distance and 500 $\mu$m wide segmented electrodes in the proximity of the experimental zone. 
Two endcaps were arranged at a distance of about 10~mm and feature 1 mm through holes which allow optical access for the Rydberg excitation beam. The RF signal for trapping, generated by an amplified output of a signal generator, was impedance matched to the trap using a helical resonator \citep{siverns12a}. RF frequency ${\rm \Omega}_{\rm RF}=2\pi \times$14.56~MHz ($2\pi \times$5.98~MHz) and RF amplitude $V_{\rm {0-peak}}=250$--300~V ($V_{\rm {0-peak}}=150$--200~V) were used. Typical secular trapping frequencies were $\lbrace\omega_{r1},\omega_{r2},\omega_{z}\rbrace=2 \pi \times \lbrace 1830, 1140, 560\rbrace$~kHz. 

\begin{figure}[h!]
\resizebox{0.49 \textwidth}{!}{%
\includegraphics{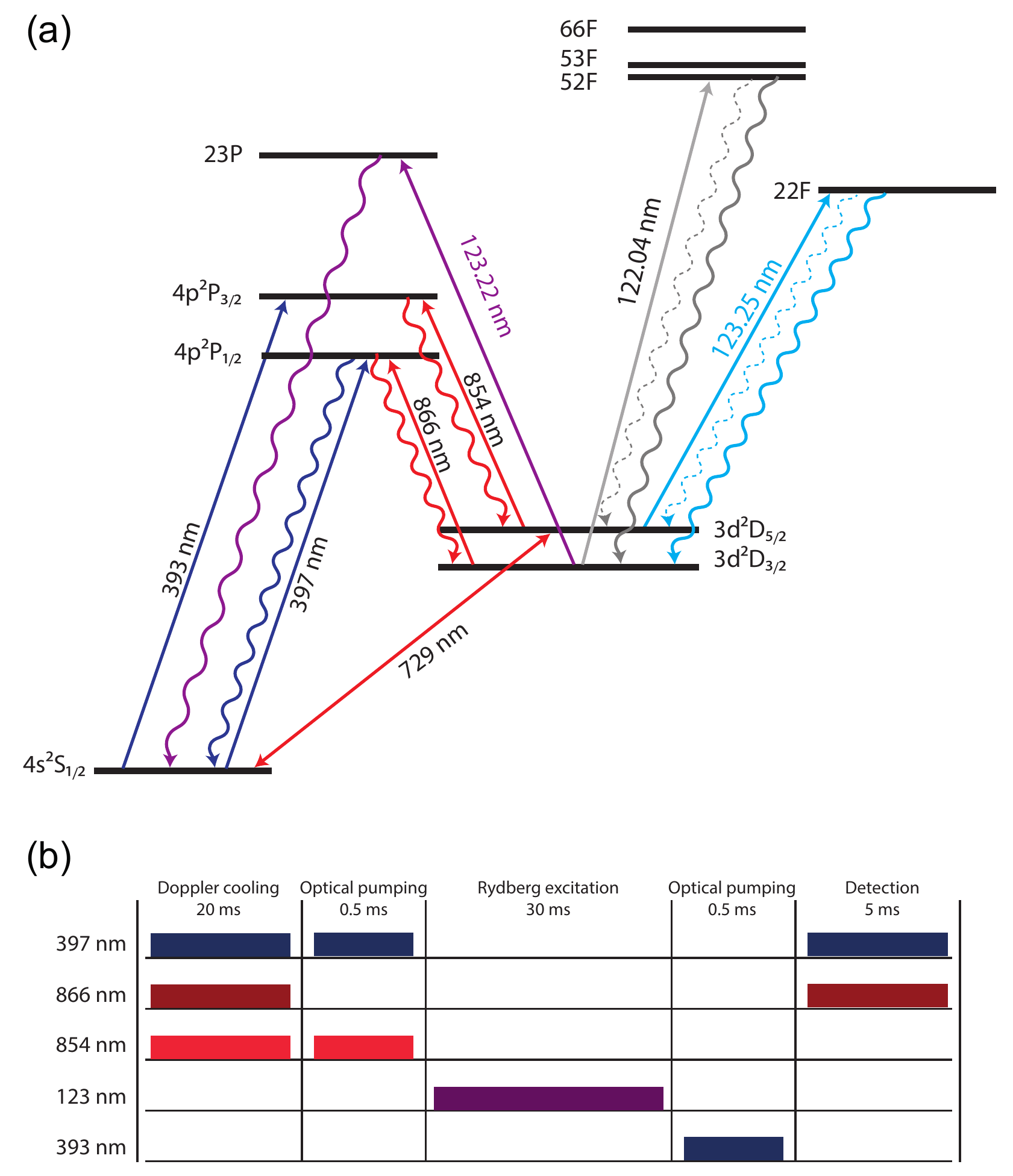}}
\caption{\label{fig:calciumElevels} (a) Reduced energy-level diagram of the $^{40}$Ca$^{+}$ ion, and relevant transitions for laser cooling, Rydberg excitation and detection. Only transitions to the Rydberg states that can be driven by our VUV laser system are shown. Only the 23\,P$_{1/2}$ line is investigated here, for the F lines, see references \citep{feldker15a, bachor16a}. (b) Laser pulse sequence for the excitation and detection of the 23\,P$_{1/2}$ state from the metastable 3\,D$_{3/2}$ state, see text for details.}
\end{figure}

To load the trap, an atomic beam of neutral calcium was produced by evaporation from a resistively heated stainless steel tube aligned towards the loading zone. The Ca beam passed through a slit to reach the trapping region where $^{40}$Ca$^+$ ions were produced by resonant photoionization using two diode laser beams near 423~nm and 375~nm. All measurements were done using single trapped ions.
Fast loading of ions has been enabled by transporting them from a separate loading zone, which was used as an ion reservoir. In this way, we mitigate parasitic electric field shifts due to surface contaminations of trap electrodes near the loading region. 
Doppler cooling of Ca$^{+}$ ions was carried out using three diode laser beams at 397, 866 and 854 nm pumping on the $(4s)^{2}$S$_{1/2}\rightarrow(4p)^{2}$P$_{1/2}$, $(3d)^{2}$D$_{3/2}\rightarrow(4p)^{2}$P$_{1/2}$ and $(3d)^{2}$D$_{5/2}\rightarrow(4p)^{2}$P$_{3/2}$ transitions, respectively. 
Sideband cooling and high-precision spectroscopy of motional modes were achieved using an ultra-stable Titanium-Sapphire laser at 729~nm that drives the $(4s)^{2}$S$_{1/2}\rightarrow(3d)^{2}$D$_{5/2}$ narrow quadrupole transition (figure~\ref{fig:calciumElevels}(a)).

Laser excitation of ions into Rydberg states is carried out using continuous-wave vacuum ultra-violet (VUV) coherent light at 122--123~nm which drives certain $n$\,P and $n$\,F transitions of the Ca$^+$ ion from the metastable 3\,D$_{3/2}$ or 3\,D$_{5/2}$ states with lifetimes in the order of 1~s~\citep{schmidtkaler11a}, see figure~\ref{fig:calciumElevels}(a).
The VUV beam enters the trap through endcap holes and passes along the trap symmetry axis. 
VUV laser radiation was generated based on a four-wave frequency mixing technique, in which three light fields at 254, 408 and 580 (555)~nm were tuned close to the 6$^{1}$S$\rightarrow$6$^{3}$P, 6$^{3}$P$\rightarrow$7$^{1}$S and 7$^{1}$S$\rightarrow$10$^{1}$P (11$^{1}$P) transitions in mercury to enhance the efficiency of the non-linear process~\citep{eikema99a,kolbe12a,bachor16a,feldker16b, schmidtkaler11a, bachor18a}. 
Having the three fundamental beams locked to a reference cavity using the Pound-Drever-Hall technique \citep{drever83a}, we estimated the VUV laser bandwidth of 850(130)~kHz \citep{bachor18a}. 
The wavelength of the fundamental beams were monitored by a wavelength meter (High Finesse WSU-10), which is calibrated to the 4\,S$_{1/2}\rightarrow$~3\,D$_{5/2}$ quadrupole transition of $^{40}$Ca$^+$, and is accurate to about 10 MHz at the sum frequency.
The VUV beam was collimated and focused to the trap experimental zone in an ultra-high vacuum chamber (at pressure $<5\times10^{-10}$~mbar) and the beam intensity is monitored by a photomultiplier tube. The efficiency of the four-wave-mixing process and thus the beam intensity is a sensitive function of the wavelength generated~\citep{schmidtkaler11a}. For the presented measurements, the laser intensity is about $3.3 \times 10^{3}$~W/m$^{2}$ at 123.217~nm near the trap centre. We estimate a VUV beam waist of about 12~$\mu$m and power of 1.5~$\mu$W at the position of ions.
 
A successful excitation to a Rydberg state is detected from an electron shelving signal~\citep{feldker15a} resulting in either the ion transferred into the 3\,D$_{5/2}$ state, thus no-fluorescence, or the ion undergoing fluorescence cycles on the 4\,S$_{1/2}$ to 4\,P$_{1/2}$ transitions when exposed to laser light near 397~nm and 866~nm. The fluorescence light near 397~nm  was collected, and spatially resolved by a microscope (numerical aperture~$=0.27$) and was imaged onto an electron-multiplying charge-coupled device (EMCCD) camera.

The laser pulse scheme used for the excitation and detection of the 23\,P$_{1/2}$ line is shown in figure~\ref{fig:calciumElevels}(b). A single, Doppler-cooled ion was optically pumped to the 3\,D$_{3/2}$ state from which it was excited to the 23\,P$_{1/2}$ state. The population of the 23\,P$_{1/2}$ state decays to the 4\,S$_{1/2}$ state in multiple steps with the predicted lifetime of about 18~$\mu$s \citep{glukhov13a}. Subsequently, the ground-state population was  transferred to the 3\,D$_{5/2}$ state using the 393~nm laser beam via the short-lived 4\,P$_{3/2}$ state. Alternatively, $\pi$-pulses of the 729~nm beam that address the Zeeman manifold of the 3\,D$_{5/2}$ level can be employed. Successful Rydberg excitation events were therefore counted as dark ions in this case.

\section{Rydberg 23\,P$_{1/2}$ line shape and electric-field compensation}\label{sec:Line}
Effects of the trapping fields on Rydberg ions in a linear Paul trap can be classified into two cases; RF and static fields minima overlapped, i.e. no ``excess'' micromotion is experienced by ions, and non-overlapped fields minima, i.e. ions ``excess'' micromotion is significant \citep{berkeland98a}. 
Due to the large polarisability of Rydberg ions, ``excess'' micromotion may lead to strong driving of phonon-number-changing transitions, and to resonance frequency shifts \citep{higgins18a}, and hence, it is always preferred to work in the former regime. Minimising the axial residual RF field, which is along the Rydberg excitation beam, is of particular importance in our experiment.  

\begin{figure}
\resizebox{0.48\textwidth}{!}{%
\includegraphics{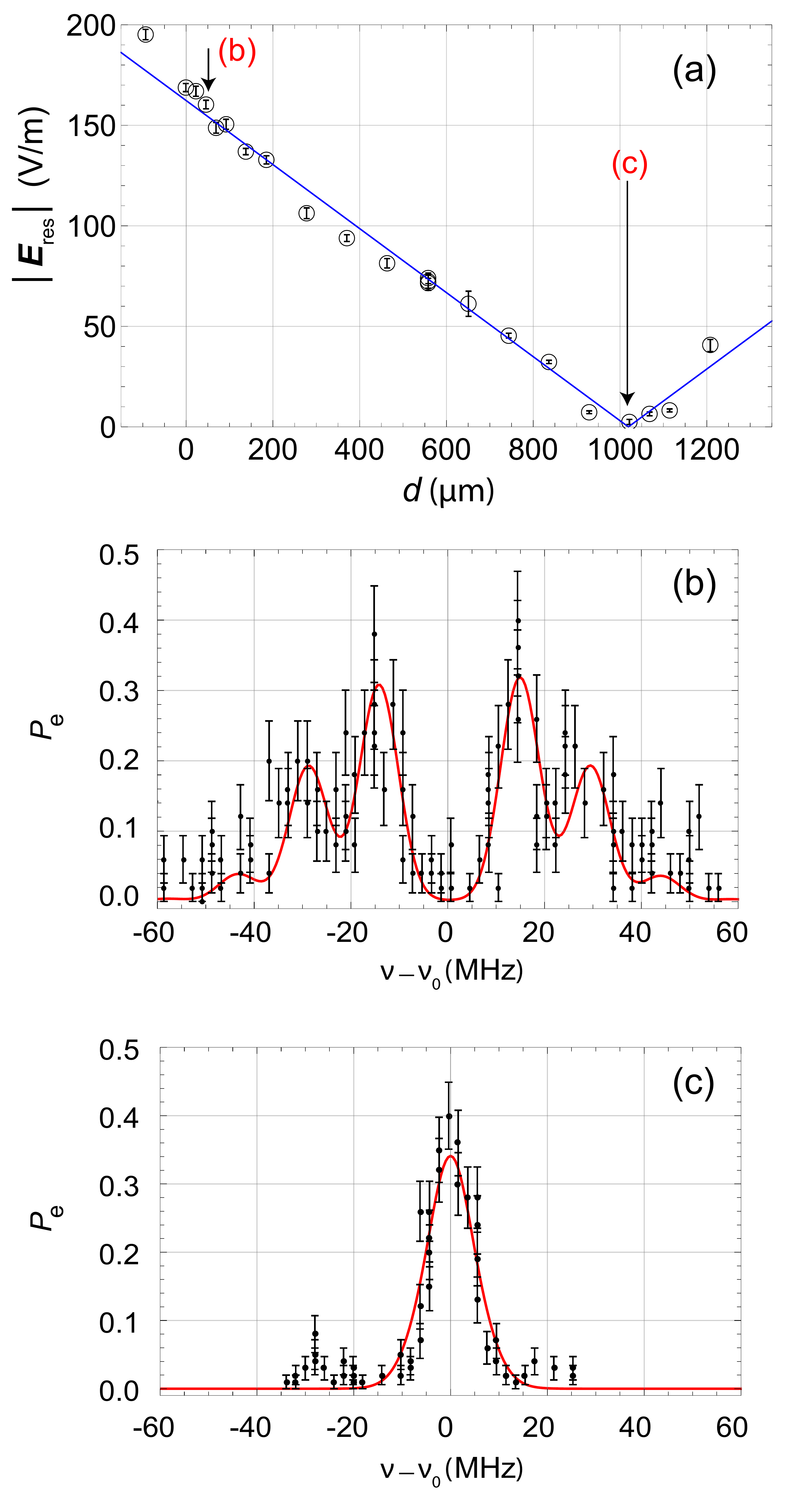}}
\caption{\label{fig:FieldCompensation_23Pline} (a) Amplitude of the residual electric field at the ion position as a function of ion displacement along the RF node. The solid line is a fit. The errorbars were calculated from fits to the Rabi oscillations on the carrier and on the first red micromotion sideband transitions, see text for details. (b) and (c) Excitation probabilities of the 3\,D$_{3/2} \rightarrow $\,23\,P$_{1/2}$ transition as a function of the VUV frequency near the resonance at the electric field $\vert \boldsymbol{E}_{\rm res} \vert=160$~V/m and $\vert \boldsymbol{E}_{\rm res} \vert <$10~V/m, respectively, as indicated in (a). The red, solid curves in (b) and (c) are calculated line shapes using the model in equation~\ref{eq:lineshape}. The errorbars depict quantum projection noise for 100 measurement cycles at each data point. In (b)(in (c)), the RF voltage $V_{\rm RF}=280$ (120)~V and the RF frequency ${\rm \Omega}_{\rm RF}$=$2\pi\times$ 14.56 (5.98)~MHz were applied. Note that (a) shows the result obtained from micromotion spectroscopy on a narrow transition using a 729~nm laser, however; (b) and (c) are the Rydberg lines measured using VUV radiation at 123.217~nm.}
\end{figure}

Resolved sideband spectroscopy on the $(4s)^{2}$S$_{1/2}(m_{j}=-1/2)\rightarrow(3d)^{2}$D$_{5/2}(m_{j}=-5/2)$ transition enabled us to precisely determine the RF electric field amplitude $\vert \boldsymbol{E}_{\rm res}\vert$ at the ion position.
If the ion is exposed to an oscillating electric field, this narrow quadrupole transition acquires sidebands. We measured the excitation strength as a Rabi frequency of the carrier and the first micromotion sidebands from which the modulation index due to the micromotion, and hence, the electric field amplitude were determined~\citep{roos2000a}.     
Figure~\ref{fig:FieldCompensation_23Pline}(a) shows the result of this measurement, where a single ion was moved to characterise the electric field along the line of the RF node. 
At each position, when moving the ion along the RF node, we minimised micromotion in the transversal plane perpendicular to the trap axis by applying compensation voltages on trap electrodes. To detect such radial micromotion, we observe the micromotion-induced modulation of the fluorescence intensity from the 4\,S$_{1/2}$ to 4\,P$_{1/2}$ transition~\citep{berkeland98a}.
For the radial field compensation, we estimated that the method is accurate to about 4 to 8~V/m for different trapping frequencies, and hence, we estimate a slightly better sensitivity for the field compensation in the radial directions as compared to the axial one. 
Our measurement shown in figure~2(a) indicates that the axial component of the micromotion is minimised at about 1000~$\mu$m from the trap geometry centre. We conjecture that this effect might arise due to imperfections of the trap geometry.   
All further spectroscopic measurements for the transition frequency have been carried out at this position, using a single ion while stray electric fields along radial and axial directions were compensated.

From a Gaussian fit to the Rydberg excitation probabilities that have been measured as a function of the frequency of the 123 nm beam, we have determined the 3\,D$_{3/2} \rightarrow $\,23\,P$_{1/2}$ absolute, measured frequency at 2\,433\,043\,773(12)~MHz.
Systematic frequency shifts of Rydberg transitions of trapped ions can be caused by the ac Stark shift \citep{higgins17a}. This shift quadratically depends on the polarisability of Rydberg states, which scales with $n^{7}$ for hydrogen-like ions, where $n$ is the principal quantum number~\citep{kamenski14a}. Using the predicted polarisability of about $\mathcal{P}_{23 \rm P}\approx-1.2\times 10^{-32}$~C$^2 \cdot$m$^2 \cdot$J$^{-1}$ for the 23\,P$_{1/2}$ state~\citep{kamenski14a}, we estimated less than $-1.4$~kHz frequency shift at 10~V/m, a conservative value for the residual electric field at the ion position. The ac Stark shift for the line shown in Fig 2(b), where the ion is subjected to $\vert \boldsymbol{E}_{\rm res} \vert =160$~V/m is about $-22$~kHz.
 
A second systematic shift arises from the ion thermal oscillation. The large polarisability of Rydberg states leads to the quadratic ac Stark shift that modifies motional frequencies of the ion in a harmonic trap. Therefore, the energy for Rydberg excitation is shifted depending on the phonon numbers in normal modes of oscillation. Since such a trap frequency alternation is two orders of magnitude larger for radial modes as compared to an axial mode in a typical experiment, the shift of the Rydberg line transition is mainly affected by radial phonon distributions. This shift depends on the sign of the polarisability of Rydberg states, e.g., shifts towards larger frequencies for $n$\,P states of Ca+ with negative polarisabilities (see measured shifts towards smaller frequencies for Rydberg S states of Sr$^{+}$ with positive polarisabilities ~\citep{higgins18a}). Doppler cooling limit in this case is estimated to be about 5~mK. For our typical operation conditions and a Doppler-cooled ion, we estimated about 2~MHz frequency shift.

The observed linewidth is about 12~MHz (FWHM). 
The natural linewidth of the Rydberg levels, which scales with $n^{-3}$ \citep{kamenski14a}, is less than 10~kHz for the 23\,P line of Ca~II. 
The ac Stark broadening is negligible in this case because of low susceptibility of this P line.
For the 3\,D$_{3/2} \rightarrow $\,23\,P$_{1/2}$ transition, we calculated a Rabi frequency of $2 \pi \times53$~kHz using the 123~nm beam parameters given in section~\ref{sec:Exp}, and the light shift of about $330$~kHz.
Finally, the bandwidth of the excitation laser is $<1$~MHz (section~\ref{sec:Exp}). Table~\ref{table:uncertainty} summarizes an overview of the sources of systematic errors that are considered in our experiment.   
\begin{table}[h!]
\caption{Summary of the contributions to the overall uncertainty in the determination of the center frequency of the 3\,D$_{3/2} \rightarrow $\,23\,P$_{1/2}$ transition.}
\label{table:uncertainty}
\resizebox{0.94 \columnwidth}{!}{%
\renewcommand{\arraystretch}{1.4}
\begin{tabular}{p{5.6cm} p{2.0cm}}
\hline \hline
Source of error & Resulting uncertainty [MHz]\\
\hline 
Stabilisation of the VUV beam & $< 1$ \\
Calibration of the wavelength meter & \mbox{\hphantom{$11$}$10$} \\
Thermal shift due to the polarisability of the state & $\approx 2$ \\
Natural linewidth of the transition& \mbox{\hphantom{$11$}$0.01$} \\
ac Stark shift due to the trapping field & $<0.001$ \\
Light shift & $\approx 0.33$ \\
Overall uncertainty& \mbox{\hphantom{$11$}$12$} \\
\hline \hline
\end{tabular}}
\end{table}   

Figures~\ref{fig:FieldCompensation_23Pline}(b) and \ref{fig:FieldCompensation_23Pline}(c) show the excitation probabilities of the 23\,P$_{1/2}$ line measured at $\vert \boldsymbol{E}_{\rm res} \vert=160$~V/m and $\vert \boldsymbol{E}_{\rm res} \vert<$10~V/m, respectively. 
These results are in good agreement with our calculations for the line shapes in which the ac Stark shift and Doppler effect have been taken into account. The modified resonance frequency of the transition is given by \citep{feldker15a}:
\begin{equation}
\nu(t)=\nu_{0}+(\boldsymbol{k}\cdot\boldsymbol{x}_{\rm mm}{\rm \Omega_{\rm RF}} {\rm sin}{\rm \Omega_{\rm RF}}t-\frac{\mathcal{P} {\boldsymbol{E}_{\rm res} ^{2}}}{2} {\rm cos}^{2}{\rm \Omega_{\rm RF}}t)/{2\pi}.
\label{eq:freq}
\end{equation} 
Here, $\nu_{0}$ is the unaffected resonance frequency, $k$ is the wave vector of the VUV laser, ${x}_{\rm mm}$ is the ion micromotion amplitude, and $\mathcal{P}$ is the polarisability of the Rydberg state. From this equation, the alternation of the laser field seen by the ion can be derived:  
\begin{equation}
\begin{aligned}
E_{\rm laser}(t)\propto & e^{-i 2\pi \nu_{0} t} e^{i 2 \beta_{ \mathcal{P}} \rm \Omega_{\rm RF} t} \\ & \times \sum_{n}{J_{n}(\beta_{\rm mm})} e^{i n (\rm \Omega_{\rm RF} t+\pi/2)} \\ & \times \sum_{m}{J_{m}(\beta_{ \mathcal{P}}) e^{2i m (\rm \Omega_{\rm RF} t)}},
\label{eq:lineshape}
\end{aligned}
\end{equation} 
where, $\beta$ is defined as the modulation index of the Bessel function $J_{n}(\beta)$, and is given by $\beta_{\rm mm}=\boldsymbol{k}\cdot\boldsymbol{x}_{\rm mm}$ for the micromotion along the excitation beam and $\beta_{ \mathcal{P}}=\mathcal{P} {\boldsymbol{E}_{\rm res} ^{2}}/8 \rm \Omega_{\rm RF}$. 
The sidebands caused by micromotion appear at $n \times \rm \Omega_{\rm RF}$, whereas the sidebands due to the ac Stark effect occur at $2 n \times \rm \Omega_{\rm RF}$.
After compensation of the oscillating field, more than 5-fold reduction of the micromotion modulation index of these sidebands was observed.

\section{Determination of quantum defects and ionization energy}\label{sec:Qdef}
The level energy of the Rydberg 23\,P state investigated in this work and those of P and F states from previous measurements listed in table~\ref{table:Rydberglines} were fitted to the extended Ritz formula \citep{ritz1908a} 
\begin{equation}
\begin{aligned}
E_{n,l,j}=& I^{++}-\frac{Z^{2} R^{*}}{(n-\mu(E))^{2}}+ \\ &\frac{Z^{4} \alpha ^{2}R^{*}}{(n-\mu(E))^{3}}{\Big [}\frac{3}{4(n-\mu(E))}-\frac{1}{(j+1/2)}{\Big ]}.
\label{eq:1}
\end{aligned}
\end{equation}
Here, $Z$ ($=+2$) is the charge of the Ca$^{+}$ ionic core, $I^{++}$ and $\mu(E)$ denote the double ionization limit and the quantum defect, respectively. Using the recommended fundamental constants \citep{NISTasd} and the $^{40}$Ca$^{+}$ ion mass \citep{IUPACmass}, we calculated the reduced Rydberg constant $R^{*}=109$\,$735.824$\,$472$\,$7(6)$~cm$^{-1}$.  
Note that the third term in equation~(\ref{eq:1}), representing the fine-structure splitting, is significant for near-ground-state levels that are taken from reference~\citep{NISTasd} as well as for the low-lying 23\,P$_{1/2}$ and 22\,F$_{5/2}$ states.
In these calculations, we used the 3D$_{5/2}$ level energy from reference~\citep{chwalla09a} and the fine-structure splitting between the 3D$_{3/2}$ and 3D$_{5/2}$ states from reference~\citep{yamazaki08a}, which lead to uncertainties about 6 to 7 orders of magnitude smaller than those of the Rydberg states that are used in the calculations. 
The energy dependence of the quantum defect $\mu(E)$ can be approximated by a truncated Taylor expansion
\begin{equation}
\begin{aligned}
\mu(E_{n,l,j})=&\mu ^{0}_{l,j}(I^{++})+\frac {\partial \mu}{\partial E_{n,l,j}}(E_{n,l,j}-I^{++})+\\ &\frac{\partial^{2} \mu}{\partial E^{2}_{n,l,j}}(E_{n,l,j}-I^{++})^{2}+ O [(E_{n,l,j}-I^{++})^{3}]\\ &\approx \mu ^{0}_{l,j}(I^{++})-\frac {\partial \mu}{\partial E_{n,l,j}} \frac{R^{*}}{(n-\mu^{1}_{l,j})^{2}},
\label{eq:2}
\end{aligned}
\end{equation}
where $\mu ^{0}_{l,j}$, $\mu ^{1}_{l,j}$, $I^{++}$ and ${\partial \mu}/{\partial E_{n,l,j}}$ were treated as adjustable parameters in the fitting routine. All energy values $E_{n,l,j}$ were weighted by the their statistical uncertainties.
A correlated fit was constructed based on a nonlinear least squares model with 7 free parameters, while the ionization energy is used as the common fit parameter between the two series. To justify the fit results, we compared the fit residuals for fits to different data sets that are differentiated by adding or subtracting data points. For instance, we verified that fit residuals for the 5-, 6-, and 23\,P states remain small regardless to using 7 and 8\,P states or other F states listed in table~\ref{table:Rydberglines} in the fitting process.

\begin{table}[h!]
\caption{Relevant Rydberg levels of Ca~II and their energies that were used in the correlated fit to the P and F series, see text for details. For the last five terms marked with asterisks, the 3D$_{5/2}$ and 3D$_{5/2}$ energies from references~\citep{chwalla09a, yamazaki08a} were used.}
\label{table:Rydberglines}
\resizebox{0.94 \columnwidth}{!}{%
\renewcommand{\arraystretch}{1.4}
\begin{tabular}{p{2cm} p{1.9cm} p{1.6cm} p{1.6cm}}
\hline \hline
Rydberg level energy & wave number [cm$^{-1}$] & uncertainty [cm$^{-1}$]& reference(s) \\
\hline 
5\,P$_{1/2}$ &  60\,533.03 & 0.01 & \citep{NISTasd} \\
6\,P$_{1/2}$ &  74\,484.94 & 0.01 & \citep{NISTasd} \\
5\,F$_{5/2}$ &  78\,034.39 & 0.01 & \citep{NISTasd} \\
6\,F$_{5/2}$ &  83\,458.08 & 0.01 & \citep{NISTasd} \\
7\,F$_{5/2}$ &  86\,727.06 & 0.01 & \citep{NISTasd} \\
8\,F$_{5/2}$ &  88\,847.31 & 0.01 & \citep{NISTasd} \\
9\,F$_{5/2}$ &  90\,300.00 & 0.01 & \citep{NISTasd} \\
10\,F$_{5/2}$ & 91\,338.00 & 0.01 & \citep{NISTasd} \\
$^{*}$23\,P$_{1/2}$ &  94\,807.798\,8 & 0.0004 & This work \\
$^{*}$22\,F$_{5/2}$ & 94\,842.764 & 0.003 & \citep{bachor16a} \\
$^{*}$52\,F$_{5/2}$ & 95\,589.257 & 0.003 & \citep{feldker15a}\\
$^{*}$53\,F$_{5/2}$ & 95\,595.656 & 0.003 & \citep{feldker15a} \\
$^{*}$66\,F$_{5/2}$ & 95\,650.901 & 0.034 & \citep{feldker15a} \\
\hline \hline
\end{tabular}}
\end{table}   

The model compiled from equation~(\ref{eq:1})~and~(\ref{eq:2}) is usually simplified by replacing $\mu^{1}$ by $\mu^{0}$.
This form may lead to adequately good fit results, see for instance reference~\citep{deiglmayr16a} for $np$ series of Cs~I, reference~\citep{li03a} for $ns,np,nd$ series of Rb~I, and reference~\citep{lange91a} for $ns,nd,nf,ng$ series of Sr~II, in which identical results for the two cases of $\mu^{1}=\mu^{0}$ and  $\mu^{1}\neq\mu^{0}$ are verified.
This simplification however spoils the meaning of the quantum defect as discussed by Drake and Swainson \citep{drake91a}, and can ultimately limit the accuracy of the quantum defect method for precision measurements. 
In our calculations for Ca~II, we observed that the use of $\mu^{1}\neq\mu^{0}$ in the fitting procedure improved the fit ``chi-squared'' value, while fit residuals were reduced by about one order of magnitude.
This might imply that effects arising from the core penetration by the Rydberg electron~\citep{drake91a} needs to be more thoroughly investigated, for instance, the excitation of the inner-shell electrons due to the Rydberg electron can be explicitly used in multi-configuration Hartree-Fock calculations \citep{seaton83aCalcium}.

\begin{table}[h!]
\caption{Quantum defect expansion coefficients as described in equation~(\ref{eq:2}) for the $n$\,P$_{1/2}$ states of Ca~II. We extracted the values from fits to measured transitions in the given references. The uncertainties are the statistical standard deviations (one sigma) obtained from fits. The 7\,P and 8\,P terms marked with asterisks are calculated values from reference~\citep{safronova11a}. Our result, which are obtained from a correlated fit, is presented in the last row, see text for details.
}
\label{table:P}
\resizebox{1.0 \columnwidth}{!}{%
\renewcommand{\arraystretch}{1.4}
\begin{tabular}{p{3.4cm} | p{1.5cm} p{1.6cm} p{1.5cm}}
\hline \hline
Investigated range of $n$ & $\mu ^{0}$&$\partial \mu / \partial E$ [$R^{*-1}$]& $\mu ^{1}$ \\
\hline 
$32$ to $95$ \citep{cbxu98a}&$1.43(4)$&\mbox{\hphantom{$-$}$0.3(1.8)$}&$33(18)$\\
$5$, $6$, $7^{*}$, $8^{*}$ \citep{NISTasd, safronova11a}&$1.438(2)$&$-0.10(3)$&\mbox{\hphantom{$-$}$1.5(0.4)$}\\
$5$, $6$, $23$ (this work)&$1.43690(3)$&$-0.1036(4)$&\mbox{\hphantom{$-$}$1.438(4)$}\\
\hline \hline
\end{tabular}}
\end{table}

\begin{table}[h!]
\caption{Quantum defect expansion coefficients as described in equation~(\ref{eq:2}) for the $n$\,F$_{5/2}$ states of Ca~II. The uncertainties are the statistical standard deviations (one sigma) from fits.  
}
\label{table:f}
\resizebox{1.0 \columnwidth}{!}{%
\renewcommand{\arraystretch}{1.4}
\begin{tabular}{p{3.4cm} | p{1.5cm} p{1.6cm} p{1.5cm}}
\hline \hline
Investigated range of $n$ & $\mu ^{0}$& $\partial \mu / \partial E$ [$R^{*-1}$] & $\mu ^{1}$ \\
\hline 
$5$ to $10$ \citep{NISTasd}&$0.02930(8)$&$0.040(2)$&$0.14(7)$\\
$5$ to $10$, $22$, $52$, $53$, $66$ (this work)&$0.02902(2)$& $0.0332(4)$ &$0.46(2)$\\
\hline \hline
\end{tabular}}
\end{table}

Table~\ref{table:P}~and~\ref{table:f} present the results for quantum defects of the P and F states of Ca~II. Also listed are the values that we extracted from fits to the most precise measured data prior to the present work. 
For the P-series quantum defect, we have compared our results to the experimental data reported in reference~\citep{cbxu98a}. 
In this case, we calculated the P-series quantum defect for the centre of gravity of the $n$\,P$_{1/2}$ and $n$\,P$_{3/2}$ states, since the fine-structure splitting was not resolved in that measurement. 
We exclude the values of reference~\citep{cbxu98a} for the following reason. If we take the ionization energy from reference~\citep{cbxu98a} and the quantum defect from the correlated fit to our experimental data, we would get 120~GHz deviations of the predicted Rydberg energies from the measured values of the P and F states in this work and in references~\citep{feldker15a,bachor16a}.
This deviation is about two orders of magnitude larger than the uncertainties reported. 
For the F quantum defect, we compared our results with those extracted from solely the term energies from reference~\citep{NISTasd}.
A theoretical prediction of the values given in Tables 3 and 4 can be found in reference~\citep{djerad91a}.
 
\begin{figure}
\resizebox{0.48\textwidth}{!}{%
\includegraphics{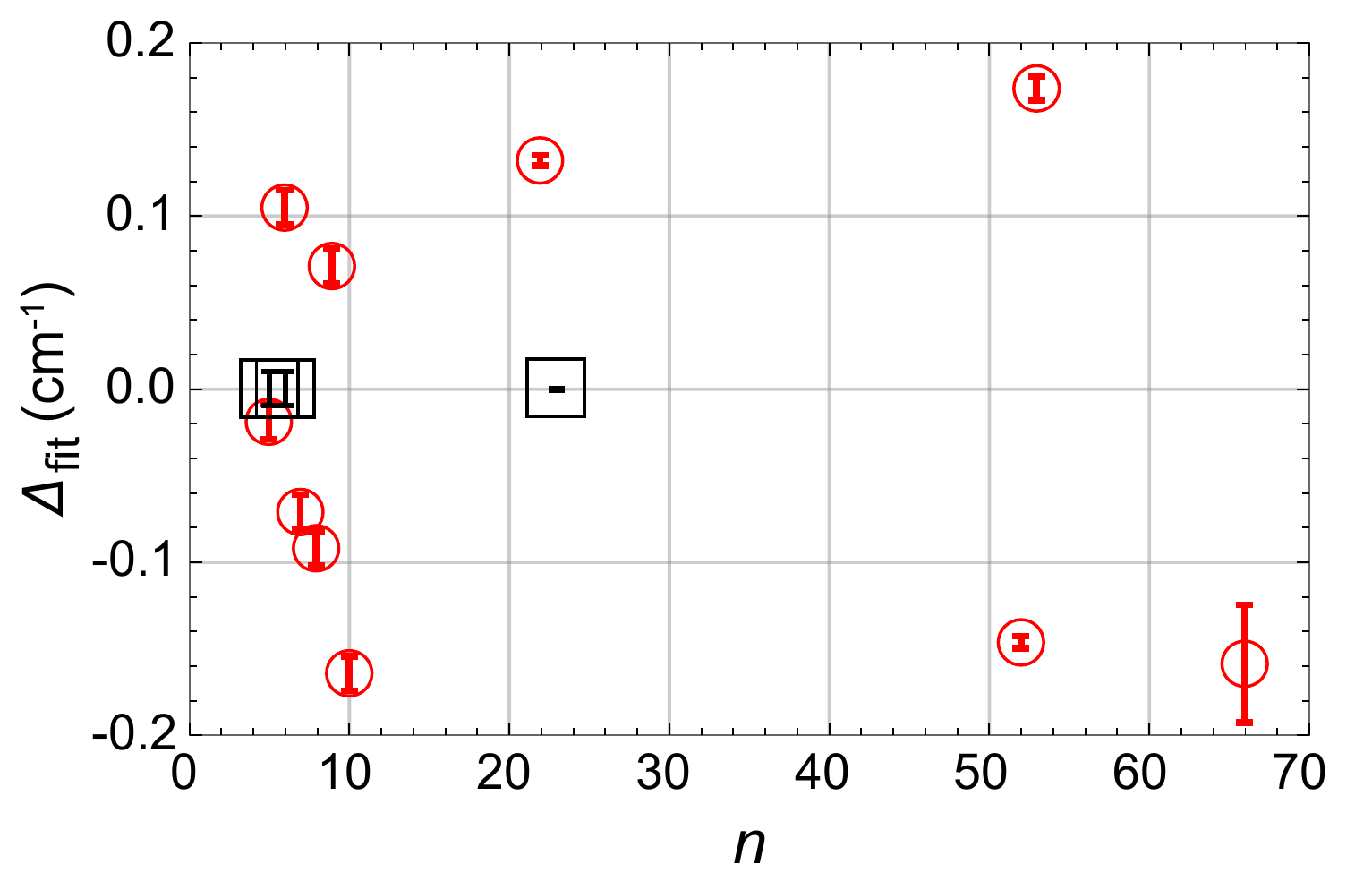}}
\caption{\label{fig:fitresidulas} Fit residuals of the correlated fit to the $n$\,P (black squares) and $n$\,F (red circles) states. Data points from references~\citep{feldker15a, bachor16a, NISTasd} are used in combination with the 23\,P transition energy reported here, see table~\ref{table:Rydberglines} and text for details.}
\end{figure}

Figure~\ref{fig:fitresidulas} shows the residuals of the correlated fit. 
The residuals for the P states agree all with zero, however, there is a discrepancy for the F states that were taken from three references~\citep{feldker15a, bachor16a, NISTasd}. For the F states, we conjecture that the scattered data in case of excitation to F states stems from unknown systematic errors for these separate sets of measurements. The apparent regularity for the P-level quantum defect could possibly signal overfitting. A full set of consistent measurements in future will allow a more accurate determination of quantum defect parameters. 


The ionization energy for the $^{40}$Ca$^{+}$ ion obtained from the correlated fit is $95$\,$751.916(32)$ cm$^{-1}$. The quoted uncertainty was calculated by adding the statistical standard deviation from the fit and the systematic errors described in section~\ref{sec:Line}.
A comparison between the ionization limits of Ca~II that have been reported since 1925 is presented in table~\ref{table:ion}. 
Our value for this quantity is consistent with the accepted value since 1999, by Litz\'{e}n et al.~\citep{NISTasd}.
\begin{table*}[ht!]
\centering
\caption{Ionization energy of Ca~II and applied measurement methods in chronological order over the last 100 years. In each row, the original reference appears without asterisk, whereas further classical references which referred to that original work are marked with asterisks. 
}
\renewcommand{\arraystretch}{1.0}
\label{table:ion}
\begin{tabular}{p{0.07\linewidth}p{0.14\linewidth}p{0.36\linewidth}p{0.14\linewidth}}
\hline \hline  
Year & $I^{++}$ [cm$^{-1}$] & Method & reference(s) \\
\hline \hline 
1925 & $95$\,$748.0$ & Absorption spectroscopy (Hilger apparatus) & \citep{saunders25a}, \citep{moore49a}$^{*}$ \\
1956 & $95$\,$751.87(3)$ & Analysis of Rydberg G series ($n=5$ to $9$) & \citep{edlen56a}, \citep{sugar85a}$^{*}$ \\
1985 & $95$\,$751.9$ & Analysis of low-lying lines (3\,D to 5\,P) & \citep{radzig85a} \\
1998 & $95$\,$748.0(.4)$ & Analysis of Rydberg P series~($n=32$ to $95$) & \citep{cbxu98a} \\
1999 & $95$\,$751.88(1)$ & Analysis of collective data, improvement by 4$p^{2}$P$^{\rm o}$ levels measurement & \citep{NISTasd}, \citep{morton03a}$^{*}$, \citep{sansonetti05a}$^{*}$ \\
2018 & $95$\,$751.916(32)$ & Analysis of Rydberg P~($n=5$, $6$, $23$) and F~($n=5$ to $10$, and $22$, $52$, $53$, $66$) levels & This work \\ 
\hline \hline
\end{tabular}
\end{table*}

\section{Summary}\label{sec:Conclusion}
We have studied the excitation of Doppler-cooled Ca$^{+}$ ions in a RF trap the 23\,P$_{1/2}$ state using VUV radiation near 123.217~nm. The modulation of this transition due to the residual RF trapping field was understood and this effect was minimised. The measured term energy was used to determine the quantum defect for the $n$\,P$_{1/2}$ states.
Using this result for the P-series quantum defect, one can narrow down the frequency range for searching the excitation energies to other P levels.
The ionization energy was determined to be 95751.916(32)~cm$^{-1}$, which is in agreement with the accepted value~\citep{NISTasd}.

\hfill \break
A. M. acknowledges the funding from the European Union's Horizon 2020 research and innovation programme under the Marie Sk{\l{}}odowska-Curie grant agreement No.~796866 (Rydion).
We acknowledge additional funding from DFG SPP 1929 “Giant interactions in Ryd-berg~Systems” (GiRyd) as well as the ERA-Net QuantERA for the project ERyQSenS.

\bibliographystyle{apsrev4-1}
\bibliography{myreferences}

\end{document}